\documentclass[letterpaper, 10 pt, conference]{ieeeconf}  

\IEEEoverridecommandlockouts                              

\usepackage{algorithm, algorithmic}
\usepackage{amsmath} 
\usepackage{amssymb}  
\usepackage{bm}
\usepackage{color}
\usepackage{glossaries}
\usepackage{graphicx}
\usepackage{lipsum}
\usepackage{microtype}
\usepackage{pgfplots}
\usetikzlibrary{pgfplots.groupplots}
\usetikzlibrary{shapes}
\usepackage{scalefnt}
\usepackage{subcaption}
\usepackage{tikz}
\usepackage{todonotes}
\usepackage{url}
\usepackage{wrapfig}
\usepackage{xfrac}

\glsdisablehyper

\pgfplotsset{compat=1.13}

\newacronym{acr:ard}{ARD}{automatic relevance determination}
\newacronym{acr:bo}{BO}{Bayesian optimization}
\newacronym{acr:dare}{DARE}{discrete-time algebraic Riccati equation}
\newacronym{acr:dda}{DDA}{dynamic domain adaptation}
\newacronym{acr:ei}{EI}{expected improvement}
\newacronym{acr:gp}{GP}{Gaussian process}
\newacronym{acr:lqr}{LQR}{linear quadratic regulator}
\newacronym{acr:map}{MAP}{maximum posterior}
\newacronym{acr:mboa}{MBOA}{model-based Bayesian optimization algorithm}
\newacronym{acr:mpc}{MPC}{model predictive control}
\newacronym{acr:pca}{PCA}{principal component analysis}
\newacronym{acr:pi}{PI}{probability of improvement}
\newacronym{acr:rl}{RL}{reinforcement learning}
\newacronym{acr:ucb}{UCB}{upper confidence bound}
\newacronym{acr:vd}{VD}{volume doubling}

\newcommand{\lfalgref}[1]{Alg.\,\ref{#1}}
\newcommand{\lfeqref}[1]{Eq.\,\eqref{#1}}
\newcommand{\lffigref}[1]{Fig.\,\ref{#1}}
\newcommand{\lfsecref}[1]{Sec.\,\ref{#1}}
\newcommand{\figurefontsize}{\footnotesize}
\newcommand{\dlqr}[1]{\ensuremath{\texttt{dlqr}\left(#1\right)}}
\newcommand{\norm}[1]{\left\lVert#1\right\rVert}

\definecolor{tableaublue}{rgb}{0.122,0.467,0.706}
\definecolor{tableauorange}{rgb}{1.0, 0.5, 0.055}
\definecolor{tableaugreen}{rgb}{0.172, 0.627, 0.172}
\definecolor{tableaured}{rgb}{0.839, 0.153, 0.157}

\title{\LARGE \bf
Bayesian Optimization for Policy Search \\ in High-Dimensional Systems via Automatic Domain Selection}

\author{Lukas P.\ Fr\"ohlich$^{1,2}$, Edgar D.\ Klenske$^{1}$, Christian G.\ Daniel$^{1}$ and Melanie N. Zeilinger$^{2,*}$
 \thanks{$^{1}$Bosch Center for Artificial Intelligence, Robert Bosch GmbH, 71272 Renningen, Germany. Corresponding author: {\tt\small lukas.froehlich@de.bosch.com}}%
 \thanks{$^{2}$Institute for Dynamic Systems and Control, ETH Zurich, 8092 Zurich, Switzerland}%
 \thanks{$^*$The research of Melanie N. Zeilinger was supported by the Swiss National Science Foundation under grant no. PP00P2 157601/1.}%
}

\begin{document}

\maketitle
\thispagestyle{empty}
\pagestyle{empty}

\begin{abstract}
Bayesian Optimization (BO) is an effective method for optimizing expensive-to-evaluate black-box functions with a wide range of applications for example in robotics, system design and parameter optimization.
However, scaling BO to problems with large input dimensions (\textgreater 10) remains an open challenge.
In this paper, we propose to leverage results from optimal control to scale BO to higher dimensional control tasks and to reduce the need for manually selecting the optimization domain.
The contributions of this paper are twofold:
1) We show how we can make use of a learned dynamics model in combination with a model-based controller to simplify the BO problem by focusing onto the most relevant regions of the optimization domain.
2) Based on (1) we present a method to find an embedding in parameter space that reduces the effective dimensionality of the optimization problem.
To evaluate the effectiveness of the proposed approach, we present an experimental evaluation on real hardware, as well as simulated tasks including a 48-dimensional policy for a quadcopter.
\end{abstract}

\section{INTRODUCTION}

\gls{acr:bo} is a powerful method for the optimization of black-box functions which are costly to evaluate.
One of the great advantages is its sample efficiency, enabling a wide variety of applications, ranging from hyperparameter search for machine learning algorithms \cite{Snoek2012PracticalBO}, medical applications \cite{Gonzalez2014GeneDesignBOWorkshop}, to robotics \cite{Calandra2016GaitOptimization, Martinez2007ActivePolicyLearning}.
Especially for high-dimensional problems, however, the number of function evaluations---experiments on the hardware in many cases---can still be prohibitive.
In this paper, we consider \gls{acr:bo} in the context of direct policy search for systems with continuous state/action space.
This offers the possibility of exploiting knowledge about the problem, and rather than considering the objective function as a black-box, we take a gray-box approach.

For many high-dimensional objective functions it is valid to assume some underlying structure, alleviating the curse of dimensionality.
One common assumption is that the effective dimensionality of the objective function is lower and lies in a linear subspace of the parameter domain \cite{Wang2016Rembo}, \cite{Garnett2014activeEmbeedingLearning}.
Another assumption is that the objective has an additive structure and many parameters are uncorrelated \cite{Kandasamy2015HighDimBO}.
However, finding an appropriate subspace is hard and the effective dimensionality of the objective function is usually not known a-priori.

\begin{figure}[t]
  \newcommand{\markerglobaloptimum}{\raisebox{0.5pt}{\tikz{\node[draw,scale=0.3,regular polygon, regular polygon sides=3,tableaured, fill=tableaured,rotate=0](){};}}}
  \newcommand{\markerestimatedoptimum}{\raisebox{0.5pt}{\tikz{\node[draw,scale=0.4,circle,tableaugreen,fill=tableaugreen](){};}}}
 \centering
 {\figurefontsize\input{figures/tikz/2d_illustrative_example.tex}}
 \caption{Illustrative 2D example depicting different optimization domains (see \lfsecref{sec:domain_selection_for_bo}) and the domains' growth due to \gls{acr:dda} (see \lfsecref{sec:dynamic_domain_adaptation}), as well as the evaluated points  ({\color{tableaublue}$\bm{\times}$}). When an estimated optimum (\protect\markerestimatedoptimum) is on the domain's boundary ({\color{black} \rule[1.25 pt]{10 pt}{2pt}}), the domain grows in the respective direction. The global optimum is marked by \protect\markerglobaloptimum.}
 \label{fig:2d_example_eigenspace_comparison}
\end{figure}

In addition to the problem of finding a suitable embedding for dimensionality reduction, it is not clear how to set the domain boundaries in which the parameters are to be optimized in a meaningful manner.
Oftentimes, these boundaries have to be found experimentally or tuned by a domain expert, introducing many additional parameters that significantly influence the convergence of \gls{acr:bo}.
We address both issues by using a learned dynamics model in combination with a model-based technique from optimal control.

The idea of combining \gls{acr:bo} and methods from optimal control has previously been explored in order to efficiently tune policies for dynamical systems \cite{Marco2016LQRBO, Bansal2017DynamicsLearningBO}.
However, both of these approaches are susceptible to model bias because the respective policy parameterizations depend on a dynamics model.
The proposed method in this paper uses a model-based approach only for selecting an appropriate parameter space and initial domain boundaries.
The optimization of the policy itself is done in a model-free manner, allowing for higher flexibility and improved final performance.

In particular, our contributions are as follows, see also \lffigref{fig:2d_example_eigenspace_comparison}:
1)~Using a learned dynamics model, we show how to automatically select the boundaries of the optimization domain such that no manual tuning or expert knowledge is needed, making \gls{acr:bo} more widely applicable for the use in policy search.
2)~We show how to determine a linear embedding that exploits the structure of the objective function, thus reducing the effective dimensionality of the optimization problem.
3)~We propose a scheme to dynamically adapt the domain boundaries during optimization based on the objective function's surrogate model if the initial domain was chosen too small.
The scheme to adapt the optimization domain is not limited to the application in policy search, but can be used as a general extension to \gls{acr:bo}.
These contributions enable direct policy search based on \gls{acr:bo} for systems with significantly higher dimensionality than reported in the literature.

\section{RELATED WORK}

In recent years, \gls{acr:bo} has emerged as a promising policy search method.
In \cite{Bansal2017DynamicsLearningBO}, a stochastic optimal control problem is considered and \gls{acr:bo} is applied to optimize the entries of the linear system matrices that describe the system dynamics, which are then used to construct an \gls{acr:lqr} controller.
Similarly, \cite{Marco2016LQRBO} also uses an \gls{acr:lqr} approach, but instead of tuning the model matrices, the weight matrices of the quadratic cost function are optimized.
In contrast to the two aforementioned papers, \cite{Calandra2016GaitOptimization} directly optimizes the entries of a linear state feedback policy.
The different parameterizations of these methods are reviewed in \lfsecref{sec:linear_policy_parameterization}.

Besides linear state feedback policies, other parameterizations have been proposed in the context of \gls{acr:bo} for policy search.
Neural networks have been used in order to learn a swing-up of a double cart-pole system in simulation \cite{Frean2008NeuralNetBO}.
On average more than 200 iterations are required to learn a good policy, although no noise is present in the system.
In \cite{Antonova2017DeepKernels}, a high-fidelity simulator of a bipedal robot is used in combination with a neural network to learn a tailored kernel for the \gls{acr:gp} surrogate model in \gls{acr:bo}.
To increase the efficiency of \gls{acr:bo}, \cite{Wilson2014TrajectoryBO} proposes to use trajectory data generated during experiments by using a so-called behavior-based kernel, which compares policies by the similarity of their resulting trajectories on the system.
However, this approach is limited to stochastic policies.
In all methods discussed above, the domain over which the parameters are optimized is always chosen manually, indicating expert knowledge or experience from other experiments.
The method proposed in this paper, in contrast, reduces the need for prior information and manual tuning.

One of the most prominent approaches for finding a lower dimensional space in which to perform the optimization is proposed in \cite{Wang2016Rembo}.
Here, the authors assume that the effective dimensionality of the objective function is small and a randomly sampled linear mapping is used to transform the high-dimensional input parameters to the lower dimensional space.
However, the effective dimensionality of the subspace is usually not known a-priori and the choice of an appropriate optimization domain is still an issue.
Instead of randomly sampling an embedding, it has been proposed to actively learn an embedding \cite{Garnett2014activeEmbeedingLearning}.
However, the proposed method has not been used in the context of \gls{acr:bo}, but rather in the active learning setting for \gls{acr:gp} regression.
Thus, evaluation points are not selected according to optimizing an objective, instead they are selected according to their information gain w.r.t. the belief of the embedding itself.

The remainder of this paper is structured as follows:
In \lfsecref{sec:preliminaries} we formally state the policy search problem and review \gls{acr:bo}.
Furthermore, we explain the different parameterizations for linear state feedback policies that have been proposed in the literature.
In \lfsecref{sec:domain_selection_for_bo} we describe the contributions of this paper.
Results from the simulations and hardware experiments are then discussed in \lfsecref{sec:simulations_and_experiments}.


\section{PRELIMINARIES}\label{sec:preliminaries}
In this section we formally state the policy search problem and explain how \gls{acr:bo} can be employed to solve it.
Furthermore, we review different parameterizations for linear feedback policies.
\subsection{Policy Search}
Consider the problem of finding a policy, \mbox{$\pi_\theta: \mathbb{R}^{n_x} \rightarrow \mathbb{R}^{n_u}$}, mapping the state $\bm{x}$ to an input $\bm{u} = \pi_\theta(\bm{x})$, which is parameterized by \mbox{$\bm{\theta} \in \bm{\Theta} \subset \mathbb{R}^{n_\theta}$}, with the goal of minimizing a specified performance criterion, or cost function, $J$, over a fixed time horizon.
Formally, this is defined in the following optimization problem:
\begin{align}\begin{split}
\label{eq:problem_statement}
\vspace{-5pt}
\min_{\bm{\theta}}J(\bm{\theta}) & = \min_{\bm{\theta}} \sum_{t = 0}^{T}\mathbb{E} \left[ c(\bm{x}_t, \pi_\theta(\bm{x}_t))\right],\\
\text{ s.t. } & \bm{x}_{t+1} = f(\bm{x}_t, \pi_\theta(\bm{x}_t)) + \bm{\nu},
\end{split}\end{align}
where $c(\bm{x}_t, \bm{u}_t)$ is the cost of being in state $\bm{x}_t$ and applying input $\bm{u}_t$, and \mbox{$f: \mathbb{R}^{n_x} \times \mathbb{R}^{n_u} \rightarrow \mathbb{R}^{n_x}$} denotes the (generally unknown) state transition model governing the dynamics of the system that are corrupted by white noise, \mbox{$\bm{\nu} \sim \mathcal{N}(0, \bm{\Sigma}_{\nu})$}.
In this paper, we apply \gls{acr:bo} to find the parameters $\bm{\theta}^*$ of a cost-minimizing policy.

\subsection{Bayesian Optimization for Policy Search}
In \gls{acr:bo}, \gls{acr:gp} regression (see, e.g., \cite{Rasmussen2006Book}) is used to model the performance of the system, $J(\bm{\theta})$, as a function of the policy parameters $\bm{\theta}$, based on noisy observations $\hat{J}(\bm{\theta})$.
In the considered setting, one function evaluation corresponds to a rollout of the current policy  on the system, after which the new data is added: \mbox{$\mathcal{D}_{n+1} = \mathcal{D}_n \cup (\bm{\theta}_{n+1}, \hat{J}(\bm{\theta}_{n+1}))$}.
As the objective is expensive to evaluate, the goal is to only use few function evaluations to find the minimum of the cost.
After each experiment, a new evaluation point in the domain is selected by maximizing an acquisition function, $\alpha(\bm{\theta}; \mathcal{D}_n)$.
Different acquisition functions have been proposed in the literature, such as \gls{acr:pi} \cite{Kushner1964PI}, \gls{acr:ei} \cite{Mockus1975ExpectedImprovement}, and \gls{acr:ucb} \cite{Cox1992UpperConfidenceBound,Srinivas2010UpperConfidenceBound}.
They all have in common that they trade off between exploration (i.e., favoring regions of the domain where the objective function has not been evaluated yet) and exploitation (i.e., proposing the estimated optimum of the objective function).
For a thorough introduction to \gls{acr:bo}, we refer the reader to \cite{Shahriari2016BayesianOptimization}.

\subsection{Linear Policy Parameterization}\label{sec:linear_policy_parameterization}
In this paper, we consider linear state feedback policies of the form:
\begin{align}
\pi_\theta(\bm{x}) = - \bm{K}(\bm{\theta}) \bm{x}.
\end{align}
The advantage of linear policies is their low dimensionality compared with other parameterizations, such as (deep) neural networks that need large amounts of training data.
As every experiment on real hardware costs considerable amount of time (and potentially money), it is infeasible to perform hundreds or even thousands of rollouts.
Another key benefit of a linear policy is that we can leverage the relation to linear optimal controllers to increase the efficiency of \gls{acr:bo}.
Although linear policies are simple in their form, they have shown impressive results, even on complex tasks, such as locomotion \cite{Mania2018SimpleRandomSearch}.

To improve the efficiency of \gls{acr:bo}, we make use of the \gls{acr:lqr}, a method commonly applied in optimal control.
One way of approximating the problem in \lfeqref{eq:problem_statement} is to linearize the system dynamics, \mbox{$f(\bm{x}_t, \bm{u}_t) \approx \bm{A}\bm{x}_t + \bm{B} \bm{u}_t$}, and quadratize the stage cost, \mbox{$c(\bm{x}_t, \bm{u}_t) \approx \bm{x}_t^T \bm{Q} \bm{x}_t + \bm{u}_t^T \bm{R} \bm{u}_t$}.
Using these approximations, one can construct the static \gls{acr:lqr} feedback gain matrix efficiently \cite[\textsection6.1]{Stengel1986Book}), which we denote as \mbox{$\bm{K} = \dlqr{\bm{A}, \bm{B}, \bm{Q}, \bm{R}}$}.
In fact, if the true system dynamics are linear and the stage cost is quadratic, the static \gls{acr:lqr} gain matrix is optimal for the case of an infinite time horizon, i.e., $T \rightarrow \infty$.
However, it leads to suboptimal performance in the case of a nonlinear system as considered in this paper.
In the following, we review three different parameterizations for linear policies in order to apply \gls{acr:bo}-based policy search, two of which make use of the \gls{acr:lqr}.

\subsubsection{Optimizing the Gain Matrix}

In \cite{Calandra2016GaitOptimization}, \gls{acr:bo} is applied by directly optimizing the feedback gain matrix, i.e., each entry in $\bm{K}^K(\bm{\theta})$ corresponds to one optimization parameter.
Thus, the number of parameters scales linearly in the number of states and inputs.
This method is model-free, i.e., it does not make use of the \gls{acr:lqr}, and thus no (linear) dynamics model is required for this parameterization.

\subsubsection{Optimizing System Matrices}
This particular parameterization was first used in \cite{Bansal2017DynamicsLearningBO}.
The idea is to parameterize the system matrices $\bm{A}$ and $\bm{B}$, where each entry of the matrices corresponds to one parameter, from which then the respective \gls{acr:lqr} can be calculated:
\begin{align}\begin{split}\label{eq:ab_parameterization}
\bm{K}^{AB}(\bm{\theta})= \dlqr{\bm{A}(\bm{\theta}), \bm{B}(\bm{\theta}), \bm{Q}, \bm{R}}.
\end{split}\end{align}
With this parameterization, a task-specific model is learned, which can be better than, e.g., the true model linearized around an operating point.
However, the number of parameters scales quadratically with the number of states, thus making this approach infeasible for large state spaces.

\subsubsection{Optimizing Weight Matrices}
Given a linear approximation of the dynamics, this method tunes the \gls{acr:lqr}'s weight matrices instead of the system matrices \cite{Marco2016LQRBO}:
\begin{align}
\bm{K}^{QR}(\bm{\theta}) & = \dlqr{\bm{A}, \bm{B}, \bm{Q}(\bm{\theta}), \bm{R}(\bm{\theta})}.
\end{align}
Commonly, only the diagonal entries of the weight matrices are tuned, i.e., the matrices are of the following form: $\bm{Q}(\bm{\theta}) = \operatorname{diag}(10^{\theta_1}, \dots, 10^{\theta_{n_x}})$, and \mbox{$\bm{R}(\bm{\theta}) = \operatorname{diag}(10^{\theta_{n_x+1}}, \dots, 10^{\theta_{n_x + n_u}})$}, such that the number of optimization parameters is reduced to $n_x + n_u$.

For the remainder of this paper, we refer to the aforementioned methods as $K$-learning \cite{Calandra2016GaitOptimization}, $AB$-learning \cite{Bansal2017DynamicsLearningBO}, and $QR$-learning \cite{Marco2016LQRBO}, respectively.
In this paper, we propose to combine the ideas of $K$-learning and the \gls{acr:lqr} to allow for efficient optimization of high-dimensional policies.
This is achieved by selecting the initial optimization domain boundaries based on a probabilistic dynamics model and the \gls{acr:lqr}.
Additionally, the model-based approach allows us to find a linear embedding to reduce the effective dimensionality of the optimization problem as will be described in the next section.


\section{AUTOMATIC DOMAIN SELECTION \& DIMENSIONALITY REDUCTION }
\label{sec:domain_selection_for_bo}

It is well-known that \gls{acr:bo} has to cover the parameter space sufficiently well with respect to the lengthscale of the cost function in order to find a good estimate of the true optimum.
Without prior knowledge, however, it is difficult to decide on the range of the parameters for optimization, which is crucial for obtaining good performance without spending an excessive amount of function evaluations for exploration.
Commonly, the issue of finding an appropriate domain is left as tuning parameter and domains are chosen, e.g., by prior experience or problem-specific expertise.
Especially in high dimensions, manual tuning of the domain parameters is not feasible.

We address this issue and propose a technique for automatic domain selection, which consists of the following steps: first, a probabilistic model for the system dynamics is learned and, second, we then employ model-based techniques from optimal control to find a distribution over policies.
Based on this distribution we can define an appropriate domain for tuning the policy parameters using \gls{acr:bo} (\lfsecref{sec:independence_domain}).
Furthermore, we can use the distribution over policies to find an embedding that maps the policy parameters into a lower-dimensional space, thus further increasing the efficiency of the subsequent optimization (\lfsecref{sec:pca_domain}).

To obtain a probabilistic model of the system dynamics, we perform Bayesian linear regression (see, e.g., \cite[\textsection3.3]{Bishop2006Book}) using recorded data of state/action trajectories to obtain an approximate linear model of the system dynamics.
This results in a Gaussian distribution over linear dynamics models:
\begin{align}\label{eq:model_distribution}
p(\texttt{vec}(\bm{A}, \bm{B})|Data) = \mathcal{N}(\texttt{vec}(\bm{A}, \bm{B}) | \bm{\mu}^{AB}, \bm{\Sigma}^{AB}),
\end{align}
where $\bm{\mu}^{AB}$ is the \gls{acr:map} estimate, $\bm{\Sigma}^{AB}$ quantifies the distribution's uncertainty, and the $\texttt{vec}(\cdot, \cdot)$ notation denotes that the matrices $\bm{A}$ and $\bm{B}$ are reshaped and stacked to a vector.

\subsection{Independence Domain}\label{sec:independence_domain}

Based on the probabilistic model of the system dynamics, we are now seeking to define an appropriate range for all policy parameters.
From \lfeqref{eq:model_distribution} we can sample $n_s$ pairs of $(\bm{A}, \bm{B})$ for which we can each calculate the respective \gls{acr:lqr} feedback gain matrix, resulting in the sample distribution: $p(\texttt{vec}(\bm{K}) | (\bm{A}, \bm{B})_{1:n_s})$.
In general, this sample distribution can be multi-modal due to nonlinearities from the Riccati equation that is solved for the construction of the \gls{acr:lqr} \cite{Stengel1986Book}.
However, since we are only looking for a bounding box based on the samples, it is sufficient to use a unimodal approximation.
To this end, we use a product of independent normal distributions, one for each dimension:
\begin{align}
p(\texttt{vec}(\bm{K}) | (\bm{A}, \bm{B})_{1:n_s}) \approx \prod_{i=1}^{n_\theta} \mathcal{N}(\texttt{vec}(\bm{K})_i | \mu^K_i, \sigma^K_i),
\end{align}
where the individual parameters for means and variances are found using moment matching.
Given this approximation, we can construct a domain centered around the mean of the distribution, where the width is governed by the distribution's standard deviation:
\begin{align}\begin{split}
\bm{\Theta}^{K}_{\text{indep}} & = [\mu^{K}_{1} - \beta\sigma_1^{K}, \mu^{K}_{1} + \beta\sigma_1^{K}] \\
& \times \dots \times [\mu^{K}_{n_\theta} - \beta\sigma_{n_\theta}^{K}, \mu^{K}_{n_\theta} + \beta\sigma_{n_\theta}^{K}],
\end{split}
\end{align}
where the parameter $\beta$ determines the effective size of the domain.
Due to the assumption of independence between entries in $\bm{K}$, we call this domain the \textit{independence domain}.

\subsection{PCA Domain}\label{sec:pca_domain}

In the previous section, we assumed independence between policy parameters of the sample distribution.
However, in general the entries of $\bm{K}$ are not independent.
In order to take advantage of potential correlations between parameters, we can approximate the sample distribution with a multivariate Gaussian:
\begin{align}
p(\texttt{vec}(\bm{K}) | (\bm{A}, \bm{B})_{1:n_s}) \approx \mathcal{N}(\texttt{vec}(\bm{K}) | \bm{\mu}^K, \bm{\Sigma}^K),
\end{align}
where the goal of the approximation is to model the overall location and spread of the samples and not to accurately model the (potentially) multiple modes.

Now, based on the multivariate distribution, we propose to transform the optimization parameters into the eigenspace of the covariance matrix, $\bm{\Sigma}^K$.
The transformation of the parameters is then described by \mbox{$\tilde{\bm{\theta}} = \bm{T} (\bm{\theta} - \bm{\mu}^K)$}, where the transformation matrix $\bm{T}$, consists of the eigenvectors of $\bm{\Sigma}^K$.
In the eigenspace, the optimization domain is given by:
\begin{align}\begin{split}
\tilde{\bm{\Theta}}^{K}_{\text{PCA}} & = [-\beta\tilde\sigma_1^{K}, \beta\tilde\sigma_1^{K}] \times \dots \times [-\beta\tilde\sigma_{n_\theta}^{K}, \beta\tilde\sigma_{n_\theta}^{K}],
\end{split}\end{align}
where $\tilde\sigma_i^K$ denotes the $i$-th eigenvalue of $\bm\Sigma_K$.
In essence, we are performing a \gls{acr:pca} \cite[\textsection12.1]{Bishop2006Book} and hence we call this domain the \textit{PCA domain}.
The benefit of using this kind of transformation is that we 1) are able to identify the most relevant directions of the parameter space and 2) still retain the uncertainty information in each direction and thus are able to create meaningful parameter ranges in the transformed space.

For high-dimensional problems, some of the eigenvalues are often close to zero and the domain size in the respective dimension becomes negligible, effectively reducing the dimensionality of the optimization problem.
Note that the volume of the domain in eigenspace is always smaller or equal to the independence domain, thus \gls{acr:bo} in the eigenspace leads to faster convergence if there are correlations between parameters.
A visualization of the different domains is shown in \lffigref{fig:2d_example_eigenspace_comparison}.

\subsection{Dynamic Domain Adaptation}\label{sec:dynamic_domain_adaptation}

For both the independence domain and PCA domain presented in the previous sections, the tuning parameter $\beta$ needs to be chosen carefully.
With increasing $\beta$, \gls{acr:bo} has to cover a larger space during the subsequent optimization, which means that more evaluations of the cost function are needed for sufficient exploration.
At the same time, it might also be possible to find better policy parameters on a larger domain.
Thus, the goal is to find a trade-off between restricting the domain to a reasonable size and choosing a large enough domain such that we do not suffer from model bias.
We argue that it is more efficient to start with a small domain, e.g., choosing $\beta = 0.5$, and then adapting the domain boundaries during optimization to account for the fact that the global optimum might not lie within the initial domain.
In this section, we explain how to exploit the surrogate model that approximates the objective function in order to adapt the domain boundaries in a goal-oriented manner.

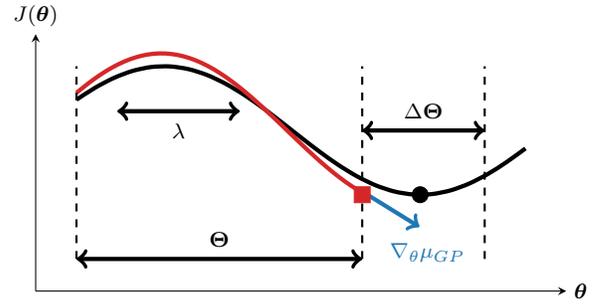
\begin{figure}
\newcommand{\markerglobaloptimum}{\raisebox{0.5pt}{\tikz{\node[draw,scale=0.4,circle,fill=black](){};}}}
\newcommand{\markerestimatedoptimum}{\raisebox{0.5pt}{\tikz{\node[draw,scale=0.4,regular polygon, regular polygon sides=4,tableaured,fill=tableaured](){};}}}
 \vspace{3mm}
 \centering
 {\figurefontsize \newcommand{\gridheight}{4}
\newcommand{\gridwidth}{6.5}
\begin{tikzpicture}
\begin{axis}[
height=5cm, width=\linewidth,
xmin=0, xmax=\gridwidth,
ymin=0, ymax=\gridheight,
axis x line = middle, 
axis y line = middle,
ticks=none,
xlabel=$\bm{\theta}$,
xlabel style={at=(current axis.right of origin), anchor=west},
ylabel={$J(\bm{\theta})$}, 
ylabel style={at=(current axis.above origin), anchor=south},
legend columns=2,
legend style={at={(0.5,-0.1)}, anchor=north, draw=none, column sep=10pt},
]
 \addplot[domain=0.5:\gridwidth-0.5, black, ultra thick,smooth] {2.5 + sin(deg(x))};
 \addplot[domain=0.5:4.0, tableaured, ultra thick,smooth] {2.5 + 1.2 * sin(1.02 * deg(x))};
 \addplot[only marks, black, mark size=3pt] coordinates{(1.5 * 3.14, 1.5)};
 \addplot[only marks, tableaured, mark=square*, mark size=3pt] coordinates{(4, 1.5)};
 \addplot[tableaublue, ->, ultra thick] coordinates{(4, 1.55) (4.7, 1.0)};
 \node[tableaublue] at (axis cs: 4.8, 0.6) {$\nabla_{\theta} \mu_{GP}$};
 \addplot[dashed, thick] coordinates{(0.5, 0.5) (0.5, 3.5)};
 \addplot[dashed, thick] coordinates{(4.0, 0.5) (4.0, 3.5)};
 \addplot[dashed, thick] coordinates{(5.5, 0.5) (5.5, 3.5)};
 \addplot[black, <->, ultra thick] coordinates{(0.5, 0.5) (4.0, 0.5)};
 \node[black] at (axis cs: 2.25, 0.8) {$\bm{\Theta}$};
 \addplot[black, <->, ultra thick] coordinates{(4.0, 2.5) (5.5, 2.5)};
 \node[black] at (axis cs: 4.75, 2.8) {$\Delta \bm{\Theta}$};
 \addplot[black, <->, ultra thick] coordinates{(1.0, 2.8) (2.5, 2.8)};
 \node[black] at (axis cs: 1.75, 2.5) {$\lambda$};
\end{axis}
\end{tikzpicture}}
 \caption{Sketch depicting the relevant parameters used for \gls{acr:dda}. The true objective ({\color{black} \rule[1.25 pt]{10 pt}{2pt}}) is approximated by a \gls{acr:gp}~({\color{tableaured} \rule[1.25 pt]{10 pt}{2pt}}) and the estimated optimum (\protect\markerestimatedoptimum) is on the domain's boundary. Consequently, the domain grows in the direction of the global optimum (\protect\markerglobaloptimum) with stepsize $\Delta \bm{\Theta}$ which is proportional to the \gls{acr:gp}'s lengthscale $\lambda$, the \gls{acr:gp}'s gradient {\color{tableaublue} $\nabla_\theta \mu_{GP}$} at the estimated optimum and the size of the current domain $\bm{\Theta}$. }
  \label{fig:1d_dda_schematic}
\end{figure}

While running \gls{acr:bo}, we have an estimate of the optimum, $\bm{\theta}^*$, i.e., the minimum of the approximate cost function on the current domain.
If \gls{acr:bo} finds the location of the estimated optimum to lie on the domain's boundary, $\partial \bm{\Theta}$, it is likely that there are better parameters outside the current domain.
Thus, we propose to grow the domain in the dimensions for which the estimated optimum is at the boundary.

\glslocalreset{acr:dda}
This \textit{\gls{acr:dda}} is guided by the \gls{acr:gp} that models the objective function.
The stepsize, $\Delta \bm\Theta_i$, by which the boundaries are increased is chosen heuristically proportional to three factors:
\begin{enumerate}
 \item The gradient of the \gls{acr:gp} posterior mean at the current estimate of the optimum, $\nabla_{\theta_i} \mu_{GP}(\bm{\theta}^*)$.
 If the gradient at the boundary is steep, we expect a potentially better point to be further away from the boundary than if the gradient is small.
 \item The lengthscale, $\lambda_i$, of the \gls{acr:gp} that approximates the cost function.
 For large lengthscales, the model assumes the cost function to vary slowly and thus the stepsize should be increased accordingly.
 \item The domain's extent in dimension $i$.
 Dimensions in which the domain is large should also be increased by a greater stepsize.
\end{enumerate}

\pagebreak
The stepsize is then given by
\begin{align}
 \Delta \bm{\Theta}_i = \gamma \cdot \nabla_{\theta_i} \mu_{GP}(\bm{\theta}^*) \cdot \lambda_i \cdot \norm{\bm{\Theta}_i},
\end{align}
where $\gamma$ is a tuning parameter that governs the effective step size.
A one-dimensional visualization of \gls{acr:dda} and all its relevant parameters can be seen in \lffigref{fig:1d_dda_schematic} and a summary of the proposed algorithm is described in \lfalgref{alg:ddabo}.
Note that \gls{acr:dda} is not limited to policy search but can be used for any \gls{acr:bo} procedure irrespective of the application and the meaning of the parameters.

A similar heuristic to grow the domain during optimization has been proposed in 
\cite{Shahriari2016UnboundedBO}, where the volume of the domain is increased isotropically by a constant factor every few evaluations of the objective function.
In the provided experiments, the constant factor is chosen to be 2, and thus this approach is called \gls{acr:vd}.
Our approach differs in two aspects: 1) the growth of the domain is not based on a fixed schedule, i.e., increasing the domain every few iterations and 2) we increase the domain anisotropically based on the surrogate model.
We argue that in the case of an initial domain that is already close the objective function's global optimum, \gls{acr:dda} only increases the domain towards the global optimum and thus leads to faster convergence compared to \gls{acr:vd}.
Besides the \gls{acr:vd} heuristic, another approach has been proposed in \cite{Shahriari2016UnboundedBO} that regularizes the acquisition function with a quadratic prior mean function.
In this way, no explicit domain boundaries need to be specified, however the shape of the quadratic regularizer gives rise to implicit bounds.
All of the aforementioned methods are evaluated on a synthetic 2D function in \lfsecref{sec:results_synthetic_functions}.

\begin{algorithm}
 \caption{\gls{acr:bo} with \gls{acr:dda} and domain selection}
 \label{alg:ddabo}
 \begin{algorithmic}[1]
  \STATE $p(\texttt{vec}(\bm{A}, \bm{B})|Data) \gets$ perform system identification to obtain probabilistic dynamics model in \lfeqref{eq:model_distribution}
  \STATE $\bm{\Theta} \gets$ select initial optimization domain, e.g., based on dynamics model (see \lfsecref{sec:domain_selection_for_bo})
  \REPEAT
  \STATE $\bm\theta^* \gets$ run BO on current domain $\bm{\Theta}$
  \IF {$\bm\theta^* \in \partial \bm\Theta$}
  \STATE $i \gets$ dimension at which $\bm\theta^*$ is at $\partial \bm\Theta$
  \STATE $\bm\Theta_i \gets $ increase domain by stepsize $\Delta \bm\Theta_i$ \\
  with $\Delta \bm\Theta_i \propto \nabla_{\theta_i} \mu_{GP}(\bm{\theta}^*)$, $\lambda_i$, $\norm{\bm{\Theta}_i}$
  \ENDIF
  \UNTIL convergence or sufficient performance is achieved
 \end{algorithmic}
\end{algorithm}

\section{SIMULATIONS AND EXPERIMENTS}\label{sec:simulations_and_experiments}

With the simulations and hardware experiments presented in this section, we aim at supporting the following results:
\begin{itemize}
 \item The methods proposed in \lfsecref{sec:domain_selection_for_bo} are able to select meaningful domain boundaries for \gls{acr:bo} and are applicable for a variety of control tasks with less manual parameter tuning.
 \item The proposed \gls{acr:dda} scheme helps to adjust the domain boundaries in a goal-oriented manner and is more efficient than the \gls{acr:vd} scheme if the initial domain is already close to the objective function's global optimum.
 \item Choosing an informed parameter transformation reduces the number of required experiments needed for good performance significantly and enables \gls{acr:bo} based policy search for higher-dimensional systems.
\end{itemize}

\subsection{Synthetic Function}\label{sec:results_synthetic_functions}
\begin{figure}
 \newcommand{\markerdomains}{\raisebox{0.5pt}{\tikz{\node[draw,scale=0.4,regular polygon, regular polygon sides=4,fill=none](){};}}}
 \newcommand{\markerglobaloptimum}{\raisebox{0.5pt}{\tikz{\node[draw,scale=0.3,regular polygon, regular polygon sides=3,tableaured, fill=tableaured,rotate=0](){};}}}
 \vspace{5pt}
 \centering
 {\figurefontsize\begin{subfigure}[b]{1.0\linewidth}
  \input{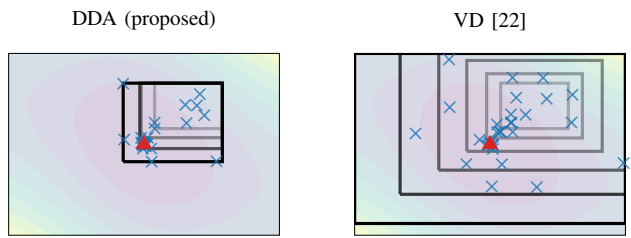}
  \caption{Parameter space view showing the growing domain boundaries~({\color{black} \rule[1.25 pt]{10 pt}{2pt}}) and evaluated points ({\color{tableaublue}$\bm{\times}$}) during optimization. The global optimum is marked by \markerglobaloptimum.}
  \end{subfigure}%
  \vspace{5pt}
  \begin{subfigure}[b]{1.0\linewidth}
   \input{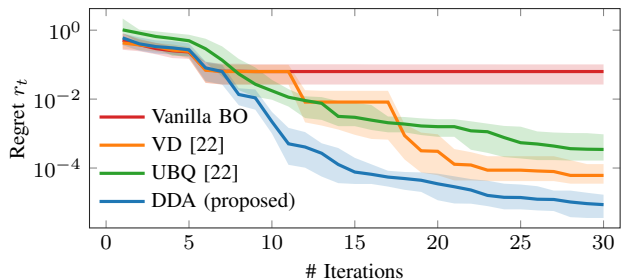}
   \caption{Regret of the best function value seen so far. With DDA the regret converges faster as the domain growth does not follow a fixed schedule (VD) and has more flexibility compared to UBQ.}
  \end{subfigure}}
  \caption{Comparison of dynamic domain adaptation (DDA), volume doubling (VD), unbounded quadratic (UBQ) and vanilla BO on the three-hump camel function.}
  \label{fig:dda_vs_vd_2d_3_hump_camel_combined}
 \end{figure}

We start by comparing the \gls{acr:dda} scheme presented in \lfsecref{sec:dynamic_domain_adaptation} with the approaches proposed in \cite{Shahriari2016UnboundedBO}.
As objective to be minimized, we choose the three-hump camel function, a 2-dimensional function with three local and one global optimum at \mbox{$\bm{\theta}^* = (0, 0)$}.
We sample $N = 50$ random initial domains of size $0.5$ in each dimension, where it was ensured that the global optimum was not contained in the initial domains.
We report the regret \mbox{$r_t = |f(\bm{\theta}^*) - \min_t f(\bm{\theta}^t)|$} for all $N$ initial domains and illustrate the domain adaptation for one of the sampled initial domains in \lffigref{fig:dda_vs_vd_2d_3_hump_camel_combined}.
The solid line represents the median over all runs and the shaded areas indicate the 25\textsuperscript{th} and 75\textsuperscript{th} percentile, respectively.
The example highlights that \gls{acr:dda} leads to improved convergence and more goal-oriented sampling of the parameter space.

\subsection{Policy Search}
\begin{wrapfigure}{R}{0.34\linewidth}
 \vspace{-20pt}
 \begin{center}
  \includegraphics[width=1.0 \linewidth]{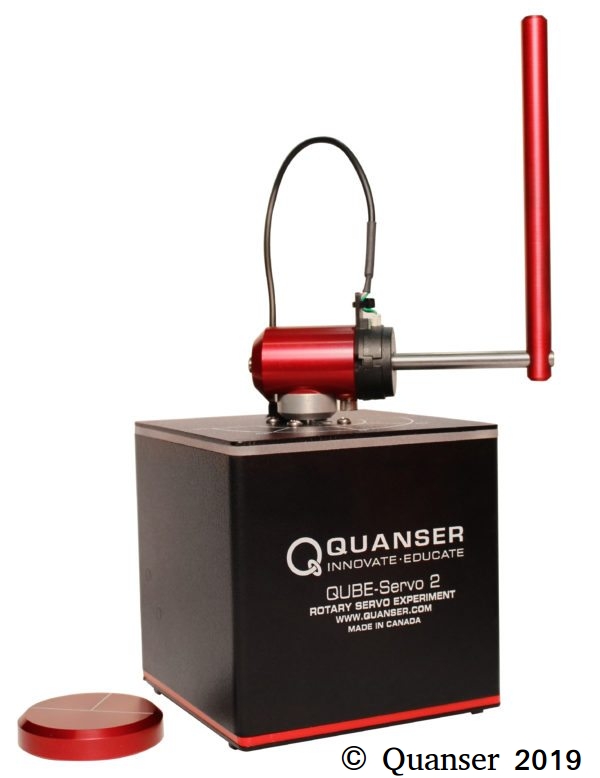}
 \end{center}
 \caption{Furuta pendulum used for hardware experiments.}
 \label{fig:quanser_qube}
\end{wrapfigure}
In this section we compare the different policy parameterizations presented in \lfsecref{sec:preliminaries} and evaluate the influence of the proposed methods in \lfsecref{sec:domain_selection_for_bo} and \lfsecref{sec:dynamic_domain_adaptation} on $K$-learning.
Additionally, we benchmark our method to REMBO \cite{Wang2016Rembo}, a well-known method to reduce the dimensionality of the objective function using random embeddings.
To evaluate performance of the policies, we compare the cost of one episode resulting from a learned policy to the cost resulting from the nominal \gls{acr:lqr} policy.
In simulation, the nominal \gls{acr:lqr} policy is based on the true dynamics model linearized at the target position and for the hardware experiments, the nominal model provided by the manufacturer is used as true model.
This results in the normalized performance measure $\eta = (J - J_{\text{LQR}}) / J_{\text{LQR}}$.
In the convergence plots shown in the following, the solid lines represent the median performance and the shaded areas indicate the 25\textsuperscript{th} and 75\textsuperscript{th} percentile, respectively.
For all experiments we use the \textit{GPyOpt}-Toolbox \cite{Gpyopt2016} and the \gls{acr:ucb} acquisition function \cite{Srinivas2010UpperConfidenceBound}, which was shown to outperform other acquisition functions in robotics applications \cite{Calandra2016GaitOptimization}.
In addition, we use the logarithm of the cost function \eqref{eq:problem_statement}, as this has been shown to lead to faster convergence \cite{Bansal2017DynamicsLearningBO}.

For the hardware experiments we use the Quanser Qube Servo 2\footnote[3]{\url{https://www.quanser.com/products/qube-servo-2/}} which has two setups: one being a disk, which is a double integrator system, the second being a Furuta pendulum \cite{Furuta1992FurutaPendulum}, see also \lffigref{fig:quanser_qube}.

\paragraph{Double Integrator (Hardware)}
\begin{figure}[]
 \newcommand{\dashedline}{\raisebox{2pt}{\tikz{\draw[-,dashed,line width = 1.5pt](0,0) -- (10pt,0);}}}
 \newcommand{\dottedline}{\raisebox{2pt}{\tikz{\draw[-,dotted,line width = 1.5pt](0,0) -- (10pt,0);}}}
 \newcommand{\bluecircle}{\raisebox{1pt}{\tikz{\node[draw,scale=0.4,circle,tableaublue,fill=tableaublue](){};}}}
 \newcommand{\orangecircle}{\raisebox{1pt}{\tikz{\node[draw,scale=0.4,circle,tableauorange,fill=tableauorange](){};}}}
 \vspace{5pt}
 \centering
 {\figurefontsize{{\input{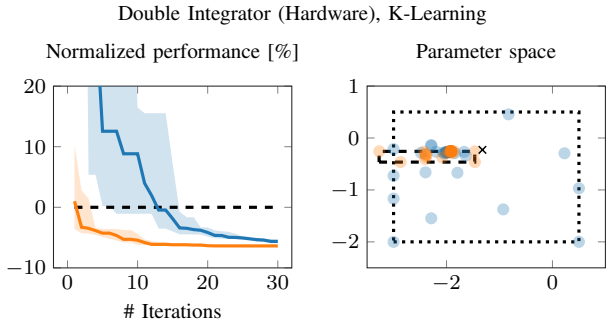}}}}
 \caption{Left: Comparison of performance when policy is optimized on the large domain ({\color{tableaublue} \rule[1.25 pt]{10 pt}{2pt}}) and the independence domain ({\color{tableauorange} \rule[1.25 pt]{10 pt}{2pt}}). 10 independent runs were performed. Right: Parameter space showing large domain (\protect\dashedline), independence domain (\protect\dottedline), evaluated policies (\protect\bluecircle /\protect\orangecircle) and  nominal LQR ($\bm{\times}$).}
 \label{fig:results_double_integrator}
\end{figure}

The state of the double integrator is 2-dimensional and consists of the disk's angle, $\phi$, and its velocity, $\dot{\phi}$.
For system identification, we applied random inputs and recorded 5 seconds of data.
Initially, the disk was placed at \mbox{$\bm{x}_0 = [\phi_0, \dot{\phi}_0]^\intercal = [90^\circ, 0]^\intercal$} with the goal of regulating it to the zero state.

In this experiment, we only apply $K$-learning, because this two-dimensional problem already shows the importance of domain selection and the domains can be easily visualized.
For a comparison between $AB$-, $QR$- and $K$-learning, we consider the dimensionality of this problem too low and thus show it on more complex systems in the following sections.

Although the problem appears simple, vanilla \gls{acr:bo} on a large domain often fails to find a sensible policy (see~\lffigref{fig:results_double_integrator}).
Optimizing on the large domain shows slow convergence as many evaluations are non-informative and far away from the optimum.
When using the proposed method in \lfsecref{sec:domain_selection_for_bo} to identify a relevant domain (independence domain in this case), \gls{acr:bo} consistently finds good policies in few iterations.
The nominal policy performs sub-optimally due to effects from friction and stiction, which are not modeled.

\paragraph{Cart-Pole (Simulation)}\label{sec:cart_pole}
\begin{figure}
 \vspace{5pt}
 \centering
 {\figurefontsize\input{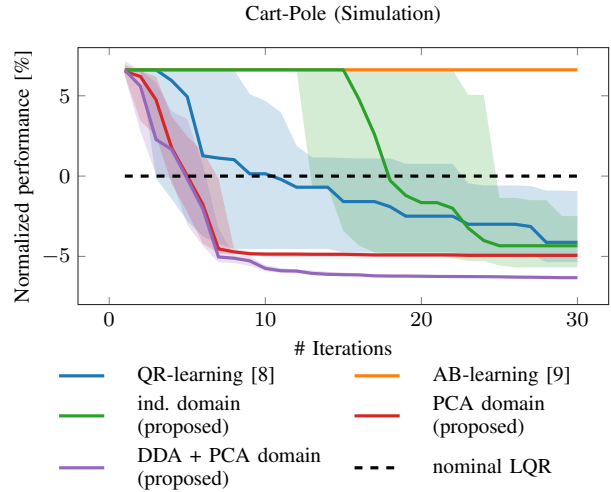}}
 \caption{Comparison of our proposed extensions to $K$-learning with $QR$- and $AB$-learning on the cart-pole task. The dashed line corresponds to the nominal \gls{acr:lqr} using the true (linearized around the upright position) dynamics. 30 independent runs were performed.}
 \label{fig:results_cartpole}
\end{figure}
The cart-pole system consists of a cart that can be moved horizontally with a freely rotating pendulum.
The input is a horizontal force acting on the cart.
The state of the system is given by the cart's position, $z$, the pendulum's angle, $\phi$, and their respective time derivatives: $\bm{x} = [z, \phi, \dot{z}, \dot{\phi}]^\intercal$.
For system identification, we started with the pendulum in the upright position ($\phi = 0^\circ$), applied random inputs and recorded the resulting state trajectory until the absolute value of the angle was larger than 30\textdegree.
This process was repeated five times.
The task was to stabilize the pendulum at the upright position over a time horizon of five seconds, where the initial condition was set to $\bm{x}_0 = [0, 45^\circ, 2 \tfrac{m}{s}, 0]^\intercal$.
Note that with this initial condition, the system dynamics are strongly nonlinear.

In \lffigref{fig:results_cartpole} we compare $K$-learning on the independence domain with $AB$-, $QR$-learning.
Additionally, we show the improved performance of \mbox{$K$-learning} when we optimize on the \gls{acr:pca} domain, with and without using \gls{acr:dda}.
$QR$-learning and $K$-learning on the independence domain show slow convergence and high variance in the resulting performance.
$AB$-learning fails to improve within 30 iterations, which is in accordance to the results presented in \cite{Bansal2017DynamicsLearningBO}, where several hundred iterations are needed to achieve good performance for the same system and a task of similar complexity.
Only when optimizing on the \gls{acr:pca} domain using $K$-learning, we consistently outperform the \gls{acr:lqr} within only five iterations.
However, as discussed in~\lfsecref{sec:dynamic_domain_adaptation}, it is possible to converge to a sub-optimal solution when the optimization domain is chosen too conservatively, as can be seen by the red curve that converges quickly to its final performance.
When we additionally use \gls{acr:dda}, the performance is further increased and all other methods are consistently outperformed.

\paragraph{Furuta Pendulum (Hardware)}
\begin{figure}[t]
 \vspace{5pt}
 \centering
 {\figurefontsize\input{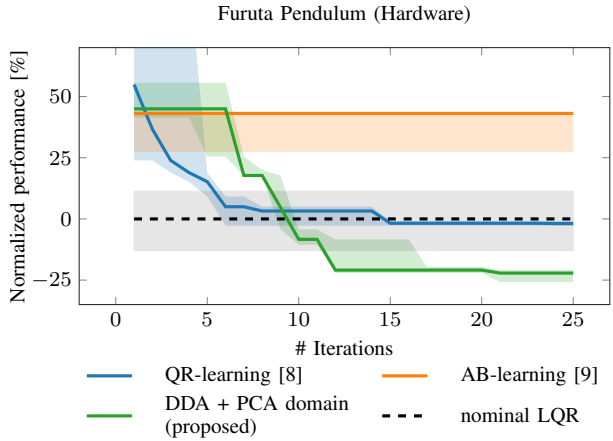}}
 \caption{Comparison of our proposed extensions to $K$-learning with $QR$- and $AB$-learning on the Furuta pendulum. 5~independent runs were performed.}
 \label{fig:results_furuta}
\end{figure}
An inverted pendulum is attached to the end of a rotary arm that is actuated via a torque; the underlying system dynamics are similar to that of a cart-pole system.
The state is given by $\bm{x} = [\eta, \phi, \dot{\eta}, \dot{\phi}]^\intercal$, with $\eta$ being the angle of the rotary arm and $\phi$ the angle of the pendulum, respectively.
The system identification process was the same as for the simulated cart-pole system and $5.7$~seconds of trajectory data were used.
The task to be optimized was regulating the system from $\bm{x}_0 = [45^\circ, 0, 0, 0]^\intercal$ to the zero state, accumulating a cost over five seconds.

Qualitatively, the results on the hardware are similar to the simulation results on the cart-pole system (see~\lffigref{fig:results_furuta})\footnote[4]{
 All experiments were initialized with the same policy, i.e., the \gls{acr:lqr} using the \gls{acr:map} estimate.
 The spread of the performance can be explained by the inherent stochasticity of the problem on the one hand and small differences in the initial conditions that are inevitable for mechanical setups on the other hand.
 Also the nominal controller shows some variance in the resulting performance, where we use the median of 20 independent runs for cost normalization.
}.
On the hardware, only $K$-learning with the proposed extensions consistently outperforms the nominal policy within $\sim$10 iterations.
The performance of $QR$-learning stagnates quickly at the nominal controller's level, which might be due to the model-based policy parameterization.
$AB$-learning is also unable to improve within the given number of iterations due to the high dimensionality of the optimization problem.
The resulting policies during $K$- and $QR$-learning result in stable behavior of the pendulum in most cases.
However, when using $AB$-learning on the hardware, we reject policies that are not in a pre-defined safe set, in order not to damage the experimental setup.
For rejected policies, we assign the same cost as obtained from the initial policy.

\paragraph{Quadcopter (Simulation)}
\begin{figure}[t]
 \vspace{5pt}
 \centering
 {\figurefontsize{\input{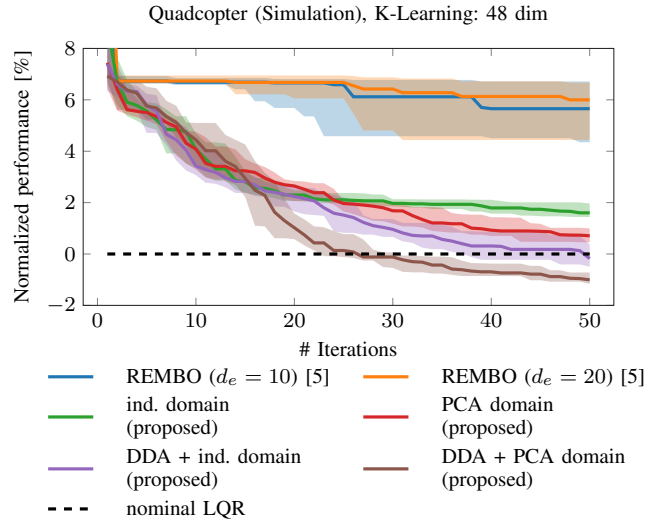}}}
 \caption{Performance comparison obtained from directly optimizing the full, 48-dimensional feedback gain matrix of a quadcopter. Our proposed extensions are compared to REMBO with two different effective dimensionalities. 30~independent runs were performed.}
 \label{fig:results_quadcopter_k_learning}
\end{figure}
To show the applicability to high-dimensional problems, we demonstrate the approach for a simulated quadcopter.
This system has twelve states (position, linear velocities, orientation and angular velocities) and four inputs (rotational velocity of the rotors).
For model learning, we recorded $50$ seconds of flying through ten different waypoints which were uniformly sampled from \mbox{$[x, y, z, \psi] \sim [-1m, 1m]^3 \times [-90^\circ, 90^\circ]$}, with $\psi$ being the yaw angle and the other states of the waypoints were set to zero.
In hardware experiments, this procedure can easily be adapted by using a remote controller to generate the data.
The control task was to stabilize the quadcopter at the hover position where the initial position was shifted $2m$ in the $x$-direction, and the roll and pitch angle were set to $30^\circ$.

Due to the superior performance of $K$-learning on previous experiments, we focus on this approach for the high-dimensional quadcopter and compare the influence of \gls{acr:dda} and the different domains as introduced in~\lfsecref{sec:domain_selection_for_bo}.
For the comparison with REMBO \cite{Wang2016Rembo}, we choose two different effective dimensionalities, $d_e = 10$ and $d_e = 20$, and optimize on the independence domain.
For all methods, we initialize \gls{acr:bo} with the \gls{acr:lqr} policy obtained from the \gls{acr:map} estimate of the system identification step, which already was able to stabilize the quadcopter.

The results are shown in \lffigref{fig:results_quadcopter_k_learning}.
With and without using \gls{acr:dda}, optimizing on the \gls{acr:pca} domain shows faster convergence in comparison with the independence domain due to the smaller volume of the domain.
Furthermore, \gls{acr:dda} helps to 1) further speed up the convergence and 2) allows for policies that consistently outperform the \gls{acr:lqr} within only 30 iterations.
Using a random embedding for dimensionality reduction as done by REMBO shows significantly lower performance and none of the optimized policies is able to outperform the nominal \gls{acr:lqr}.

\begin{figure}[]
 \vspace{5pt}
 \centering
 {\figurefontsize{{\input{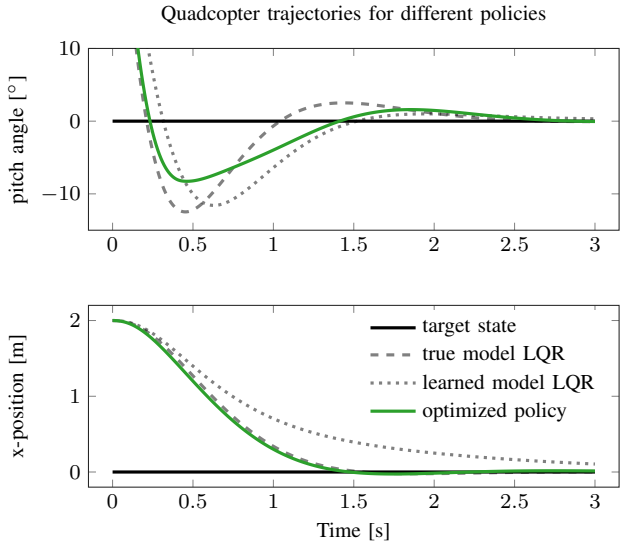}}}}
 \caption{Exemplary trajectories of the $x$-position and pitch angle before and after optimization. Note that the optimized policy leads to faster convergence to the desired states and less overshoot of the pitch angle.}
 \label{fig:results_quadcopter_trajectory_comparison}
 \vspace{-10pt}
\end{figure}

For a more intuitive understanding of the actual performance increase that is obtained with \gls{acr:bo}, we show trajectories of the quadcopter before (gray dotted) and after (green) optimization in \lffigref{fig:results_quadcopter_trajectory_comparison}.
Especially the $x$-position of the quadcopter does converge significantly faster to the desired value after optimization of the policy.

\section{CONCLUSION}\label{sec:conclusion}

In this paper, we have shown that the choice of the optimization domain is critical for the convergence and final performance, as well as scalability of \gls{acr:bo} in policy search methods.
By using a model-based technique from optimal control, we proposed an automatic domain selection method for optimizing a linear feedback policy in a model-free manner which exploits the objective functions structure and improves the sample efficiency of \gls{acr:bo}.
Additionally, we introduce a dynamic domain adaptation mechanism in order to mitigate a potential model bias due to the choice of the initial domain.
Simulations and experiments have shown that these contributions enable \gls{acr:bo}-based policy search techniques to find a policy that outperforms other control techniques that use the true dynamics model with only few system interactions.
A key benefit of the reduced search domain provided by the proposed technique is the improved scalability of \gls{acr:bo}.
We have demonstrated our approach to learn a 48-dimensional policy for a quadcopter---a size that renders standard \gls{acr:bo} techniques infeasible.

\newpage

\addtolength{\textheight}{-0cm}   


\bibliographystyle{IEEEtran}

\bibliography{bibliography/conf_names_long,bibliography/library}

\end{document}